\documentstyle[epsfig,12pt]{article}
\begin{document}
\renewcommand{\thepage}{ \arabic{page}}
\def\ab#1{\renewcommand{\theequation}{#1}}
 \def\sous#1{\addtocounter{equation}{-#1} }

\def\normal{\renewcommand{\theequation}{\arabic{equation}} }

\renewcommand{\theequation}{\thesection.\arabic{equation}}
\begin{quote}
\raggedleft
%hep-ph/yymmxxx \\
PM/97-31 \\
%\datea0
\end{quote}

\vspace*{3cm}
\begin{center}
{\bf One-Loop Minimization Conditions \\
in the Minimal Supersymmetric Standard Model 
\footnote{supported in part by EC contract CHRX-CT94-0579}}\\
\vspace{-0.3cm}
\end{center}
\setlength\baselineskip{17pt}
\begin{center}
%\vspace{-0.3cm}
\end{center}
\begin{center}
Christophe LE MOU\"EL\footnote{lemouel@lpm.univ-montp2.fr} and Gilbert MOULTAKA\footnote{moultaka@lpm.univ-montp2.fr}
\end{center}
\vspace{-0.5cm}
\begin{center}
{\it Physique Math\'ematique et Th\'eorique\\
UMR-CNRS,\\
Universit\'e Montpellier II, F34095 Montpellier Cedex 5, France}
\end{center}

\vspace{0.5cm}
\begin{abstract}

We study, in the Minimal Supersymmetric Standard Model,
the electroweak symmetry breaking conditions obtained from the one-loop 
effective potential.
Novel model-independent {\sl lower} and {\sl upper} bounds on 
$\tan \beta$, involving the other free parameters of the model, 
are inferred and determined analytically.
We discuss briefly some of the related issues and give an outlook for
further applications. 
   
.
\end{abstract}
\vspace*{2cm}
\begin{center}
{\it (extended version of the talk presented by the second author at the 
joint meeting of the European Networks `The Fundamental Structure of Matter'
and `Tests of Electroweak Symmetry Breaking',\\
Ouranoupolis, Greece, May $27^{th}$ - June $1^{st}$, 1997, to appear in the proceedings)} 
\end{center}
\newpage
\section{Introduction}

Breaking the electroweak symmetry through
radiative corrections in the MSSM \cite{EWB}, is nowadays one fashionable 
scenario. According to it,
the phenomenon of spontaneous electroweak symmetry breaking becomes intimately 
related to supersymmetry breaking.
This scenario has a double merit:
 --theoretical, as it suggests that very different
energy scales could be deeply connected through one single dynamics
--phenomenological, as it reduces tremendously the number of free parameters
of the MSSM, leading to quantitative estimates of the mass spectrum 
\cite{spectrum}.\\

Yet, the true effective potential in which the vacuum structure is encoded,
 is a poorly known object beyond the
tree-level approximation. One reason for this is the very many different mass
scales present in the MSSM, so that a renormalization group (RG) analysis 
becomes
rather tricky. Furthermore, even an RG-unimproved one-loop analysis is made
difficult by the presence of many scalar field directions.

In this talk, we report on an analytic study of the one-loop effective 
potential, and its effects on the minimization 
conditions which determine the physical electroweak vacuum \cite{nous}.
Naively, one would argue that such effects
cannot change much of the qualitative pattern of the tree-level minima.
The fact is that, even if the one-loop effective potential
is expected to differ, point-to-point, only perturbatively from its tree-level
value, it is still possible that its shape be {\sl locally} modified in such 
a way that new local minima (or at least stationary points) appear.
Furthermore, in the case of vanishing masses, the loop corrections can have
an even more drastic effect, generating perturbatively a spontaneous symmetry
breaking, the so-called dimensional transmutation \cite{CW}. The MSSM contains
by construction several free non-vanishing mass parameters, 
either of supersymmetric origin
like the $\mu$ term, or parameterizing the effective supersymmetry breaking
through the soft terms. 
Thus, the radiative electroweak symmetry breaking
(EWSB) \cite{EWB} in the MSSM does not generically realize a dimensional 
transmutation.
However, one can still sit at some energy scale where almost-flat directions 
appear\footnote{The existence of such directions is guaranteed by that of flat 
directions in the supersymmetric limit. In our case we will be 
exclusively concerned with the so-called D-flat directions}. If there is a
minimum in such a direction, then a change in the shape of the potential
at one-loop in the vicinity of the almost-flat direction, could create
{\sl perturbatively} a supplementary local minimum, which could be 
{\sl lower} than that 
already present at tree-level. The case of almost-flat direction is relevant
when $\tan \beta \sim 1$.\\

Another interesting feature has to do with the trend of the radiative 
corrections to the tree-level Higgs boson mass relations. These relations come 
up as a consequence of (softly broken) supersymmetry and of the 
{\sl minimization conditions} 
assuring the formation of vacuum expectation values associated
with $SU(2)_L \times U(1)_Y$ spontaneous symmetry breaking.
A particular property of the tree-level Higgs potential in the MSSM, 
is that the minimization  conditions
boil down to just the stationarity conditions. That is, {\sl at tree-level 
the vanishing of the $1^{st}$ order derivatives implies  that all the
squared Higgs masses are automatically positive}. 
Indeed, starting from the tree-level potential

\begin{eqnarray} 
V_{tree}&=& m_1^2 |H_1|^2 + m_2^2 |H_2|^2 + m_3^2 (H_1.H_2 +
 h.c.) \nonumber \\
&&+ \frac{g^2}{8}  ( |H_1|^2 - |H_2|^2)^2 + \frac{g_2^2}{2} 
(H_1^{\dagger}  H_2) (H_2^{\dagger}  H_1) 
\label{vtree}
\end{eqnarray}

where $m_1, m_2, m_3$ are free mass parameters,
$g^2 \equiv g_1^2 + g_2^2$,  and $g_1$ ( resp. $g_2$) denote the $U(1)_{Y}$
(resp. $SU(2)_L$ gauge couplings), and defining

\begin{eqnarray} 
<H_1>= \frac{1}{\sqrt{2}}\left( \begin{array}{c} 
               v_1 \\
               0 \\
               \end{array} \right) \hspace{2cm}
<H_2>= \frac{1}{\sqrt{2}}\left( \begin{array}{c} 
               0 \\
               v_2 \\
               \end{array} \right) 
\end{eqnarray}
where $v_1, v_2$ can always be defined to be real valued,
one ends up with four free parameters, namely, the $m_i$'s and 
$\tan \beta \equiv \frac{v_2}{v_1}$. [The combination $v_1^2 + v_2^2$ is 
assumed to be fixed by the Fermi scale]. Then, these four parameters reduce
to two when the stationarity conditions $\frac{\partial V}{\partial v_1}=
\frac{\partial V}{\partial v_2}=0$ are imposed. This is all the input
one needs to recover the celebrated Higgs boson mass relations,

\begin{equation}
m^2_{h^0, H^0}=\frac{1}{2}(m_Z^2 + m_{A^0}^2 \mp \sqrt{(m_Z^2 + m_{A^0}^2)^2 
- 4 m_Z^2 m_{A^0}^2 \cos^2 2 \beta}\;\; ) 
\end{equation}
\begin{equation}
m^2_{H^\pm} = m_{A^0}^2 + m_W^2
\end{equation}

where it is usual to choose $m_{A^0}$ and $\tan \beta$ as the two free input 
parameters. From the above relations one notes that the positivity
of the squared Higgs masses comes out {\sl automatically}, provided that
$m_A^2 \geq 0$. Furthermore, the latter is always verified, since
$m_A^2 = - (v_1^2 +  v_2^2) \frac{m_3^2}{v_1 v_2}$ and the stationarity conditions
$\frac{\partial V}{\partial v_1}=
\frac{\partial V}{\partial v_2}=0$ {\sl enforce} $v_1 v_2$ and $m_3^2$ to have
{\sl opposite} signs \cite{nous}. It should be clear that such a property
is not generally true. The question we ask here is whether it remains true
beyond the tree-level approximation. 

We will find out that the generic answer to the above question is negative.
This means that new extra conditions, on top of the ones usually imposed
in radiative EWSB scenarios, should be considered to ensure that a stationary
point is indeed a (local) minimum of the effective potential. These conditions
will imply some model independent constraints involving $\tan \beta$,
thus contrasting with the belief that such constraints are obtainable
only in model-dependent situations. Actually our constraints turn out to be even
stronger than the existing ones. Finally, the analytic study of the vacuum 
structure of the one-loop effective potential, will also allow us, on one hand,
to check whether or not new minima can occur in the regime $\tan \beta \sim 1$,
as was previously  noted, and on the other, to understand the reason for the
instability of the vacuum expectation values against the change of the
renormalization scale in the tree-level-RGE-improved approximation,
as compared to the one-loop case \cite{gamb}.  

\section{The one-loop Effective Potential}

The 1-loop effective potential has the well-known form \cite{CW}
\begin{equation}
V= V_{tree} + \frac{\hbar}{64 \pi^2} Str[ M^4 (Log \frac{M^2}{\mu_R^2} -
3/2) ] \label{EP}
\end{equation}

in the $\overline{MS}$ scheme. Here $\mu_R$ denotes the renormalization scale, 
$M^2$ the field dependent squared mass matrix of the scalar 
fields,
and $Str[...] \equiv \sum_{spin} (-1)^{2 s} (2 s + 1) (...)_s $, where the sum
runs  
over gauge boson, fermion and scalar contributions.
$V_{tree}$ is the tree-level MSSM potential \cite{susy}.
A full analytic study of the potential Eq.(\ref{EP}) in the MSSM 
is quite involved,
especially if one aims at the determination of the structure of the minima
in various scalar field directions. For instance, to keep the complete
information, the scalar, vector and fermion contributions to $M^2$
lead at best, respectively to $(14 \times 14), (3 \times 3)$ and 
$(10 \times 10)$ matrices. We have developed some analytic tools which
allow to extract information with less approximations than what is usually
done, especially in studying the first and second order derivatives of the
Logarithmic parts in Eq.(\ref{EP}), keeping simultaneously the biggest possible
number of scalar directions. In this talk we are mainly interested in the
Higgs directions in a given approximation, so we will not need all the
above mentioned developments to which we come back briefly in the last section.
  
For now, and in order to illustrate the possibility of extracting fully analytic
results, we will make a simplifying working assumption, which is 
not devoid of some physical relevance.
We will assume that the logarithms in Eq.(\ref{EP})
are re-absorbed in the running of all the parameters
in ${V}_{tree}$. This assumption would be exact,
in the context of renormalization group analysis, if there were only
one physical mass scale in the model. In our case, it amounts to the rough
approximation of neglecting altogether all field and mass scale differences
in the logs. We will also sit in the Higgs directions, that is all scalar
fields are put to zero except for the two Higgs doublets
\begin{eqnarray} 
H_1= \left( \begin{array}{c} 
               H_1^0 \\
               H_1^{-} \\
               \end{array} \right) \hspace{2cm}
H_2= \left( \begin{array}{c} 
               H_2^{+} \\
               H_2^0 \\
               \end{array} \right) 
\label{Hdoublets}
\end{eqnarray}
 
The effective potential takes then the form
\begin{equation}
 V= \overline{V}_{tree}(\mu_R^2) + \frac{\hbar}{64 \pi^2} (-3/2) Str M^4 
\label{Poteff}
\end{equation}

where $\overline{V}_{tree}(\mu_R^2)$, obtained from ${V}_{tree}$ given in
Eq.(\ref{vtree}) by the 
replacements $m_i^2 \to \overline{m_i}^2(\mu_R^2), 
g_i \to \overline{g_i}(\mu_R^2) $, is the so-called  RGE-improved
tree-level effective potential\footnote{In principle one should also
take into account the running of the Higgs fields, however the corresponding
effect is negligibly small at the energy scale we consider. 
Note also that the other running scalar fields remain consistently vanishing
if their initial value is zero}. 
The above rough assumptions about the Logs
can be improved by switching on more and more different mass scales, 
and treating separately different regions of the scalar field space 
\cite{multiscale}. At each energy scale, the one-loop effective potential
would still have the form of Eq.(\ref{Poteff}) except that its validity
would be limited to a given region of the scalar field space, and accordingly
the running of the various parameters absorbs the decoupling effects of 
higher mass scales. It follows,
since we are interested here mainly in the interplay between
the finite (non-logarithmic) terms and the functional
form of $\overline{V}_{tree}(\mu_R^2)$, that our approximation in 
Eq.(\ref{Poteff}) should be reasonably illustrative.\\

\section{The one-loop Minimization Conditions}
Putting together the contributions of the full-fledged MSSM, we find
the following form for the one-loop effective potential in the Higgs
directions (and in the Landau gauge), 
\begin{eqnarray}
V_{tree} +\kappa Str M^4&=& X_{m_1}^2 |H_1|^2 + X_{m_2}^2 |H_2|^2 + 
          X_{m_3}^2 (H_1.H_2 + h.c.)  \nonumber\\
         &&+ X ( |H_1|^2 - |H_2|^2)^2 +
        \tilde{\beta} |H_1^{\dagger}  H_2|^2 
+ \tilde{\alpha} (|H_1|^4 - |H_2|^4) + \Omega_0  \nonumber\\
\label{vloop}
\end{eqnarray}
where
$H_1.H_2 \equiv \epsilon_{i j} H_1^i H_2^j$
and
\begin{eqnarray}
\tilde{\alpha} &=& \frac{3}{2} \kappa g^2({Y_t}^2 - {Y_b}^2) \label{alpha}\\
\tilde{\beta}&=& \frac{g_2^2}{2} + \kappa g_2^2 ( g_1^2 + 5 g_2^2 - 6(Y_t^2 +
Y_b^2) ) \\
X &=& \frac{g^2}{8} + \kappa (g_1^2 g_2^2  +
   \frac{23 g_1^4 + 5 g_2^4}{4} - \frac{3}{2} g^2({Y_t}^2+ {Y_b}^2 )  \\
\nonumber
\end{eqnarray}
\begin{eqnarray}
X_{m_1}^2 &=&  m_{H_1}^2 + \mu^2 + \kappa[ -4 g_1^2 M_1^2  +
3 g_2^2 (  m_{H_1}^2 - 3 \mu^2 -4 M_2^2)\nonumber \\
&& + 12 ( Y_b^2 ( A_b^2 + m_{\tilde{b_R}}^2 +
 m_{\tilde{T}}^2 ) + \mu^2 Y_t^2)  \nonumber \\
&&+ g_1^2 ( 3 m_{H_1}^2 - 2 m_{H_2}^2 - 3 \mu^2 \nonumber \\
&&- \sum_{i=generation} 2 (
m_{\tilde{d}_R, i}^2 +  m_{\tilde{l}_R, i}^2 + m_{\tilde{Q}, i}^2 -
m_{\tilde{L}, i}^2 - 2 m_{\tilde{u}_R, i}^2) ) ] \nonumber \\
\end{eqnarray}
\begin{eqnarray}
X_{m_2}^2 &=&  m_{H_2}^2 + \mu^2 + \kappa[ -4 g_1^2 M_1^2  +
3 g_2^2 (  m_{H_2}^2 - 3 \mu^2 - 4 M_2^2) \nonumber \\
&&+ 12 ( Y_t^2 ( A_t^2 + m_{\tilde{t_R}}^2 +
 m_{\tilde{T}}^2 ) + \mu^2 Y_b^2)  \nonumber \\
&& + g_1^2 ( 3 m_{H_2}^2 - 2 m_{H_1}^2 - 3 \mu^2 \nonumber \\
&&+ \sum_{i=generation} 2 (
m_{\tilde{d}_R, i}^2 +  m_{\tilde{l}_R, i}^2 + m_{\tilde{Q}, i}^2 -
m_{\tilde{L}, i}^2 - 2 m_{\tilde{u}_R, i}^2) ) ] \nonumber\\
\end{eqnarray}
\begin{eqnarray}
X_{m_3}^2 &=& -B \mu + \kappa \mu [ g_1^2 ( B + 4 M_1 ) + 
3 g_2^2 (  B + 4 M_2) \nonumber \\
&& ~~~~~~~~~~~~~~~~~~~~~~~~~ - 12 ( A_t Y_t^2 + A_b Y_b^2)]  \\
\kappa&=&(-\frac{3}{2})\frac{\hbar}{64\pi^2} \label{kapa} \\
g^2 &\equiv& g_1^2 + g_2^2 \\
\nonumber
\end{eqnarray}

The above expressions are exact, apart from the fact that we kept only the
top/bottom Yukawa couplings $Y_t$ and $Y_b$ for simplicity, the generalization
to the other Yukawa couplings being straightforward.  
The $X_{m_i}^2$'s are functions
of the various soft susy breaking masses and couplings, associated with
all squarks doublets masses, $m_{\tilde{Q}, i}, m_{\tilde{T}}$, and singlets
masses,
$m_{\tilde{u}_R, i}, m_{\tilde{d}_R, i}, m_{\tilde{b}_R}, m_{\tilde{t}_R}$,
all sleptons doublets masses
, $m_{\tilde{L}, i}$ and singlets masses $m_{\tilde{l}_R, i}$
(no right-handed $\tilde{\nu}$), gaugino soft masses $M_1, M_2$ (no
gluino contributions at this level), Higgs soft masses $m_{H_1}, m_{H_2}$
and the supersymmetric $\mu$-term, with      
$m_1^2=m^2_{H_1}+\mu ^2$,  $m_2^2=m^2_{H_2}+\mu ^2$, 
as well as $ B \mu (\equiv -m_3^2)$ 
and the soft trilinear couplings $A_t, A_b$.
Note also that $\Omega_0$ in Eq.(\ref{vloop}) is a field independent additive 
constant
depending exclusively on soft susy breaking terms.
It contributes to the cosmological constant and will be discarded throughout
\footnote{see however \cite{multiscale1, multiscale}  for relevant issues}
We stress that no model-dependent assumptions are needed to establish the
above expressions. The results which will follow from them will thus be 
applicable either in the context of SUGRA-GUT scenarios, where eventually 
universality may be assumed \cite{hiddensector}, 
or in a gauge-mediated supersymmetry breaking context \cite{gaugemed}, 
or for that matter in any fully model-independent analysis.\\
Starting from Eq.(\ref{vloop}) we can determine both the {\sl stationarity}
conditions 
\begin{equation}
\frac{\partial V}{\partial H_a^i}{\big |}_{H_a=<H_a>} = 0
\label{station}
\end{equation}
and {\sl stability} conditions 
\begin{equation}
 \mbox{eigenvalues of} \frac{\partial^2 V}{\partial H_a^i \partial H_b^j}|_{H_a=<H_a>}
\geq 0
\label{stabili}
\end{equation}

where $a, b=1,2$ label the two Higgs doublets 
and $ i, j=1,...4$ label the four real 
valued fields within each of the doublets.
From Eq.(\ref{station}) one finds the following two conditions 
\begin{eqnarray}
X_{m_1}^2 v_1 + X_{m_3}^2 v_2 + X v_1 ( v_1^2 - v_2^2) + \tilde{\alpha} v_1^3 = 0  &,&
X_{m_2}^2 v_2 + X_{m_3}^2 v_1 + X v_2 ( v_2^2 - v_1^2) - \tilde{\alpha} v_2^3 = 0  \nonumber \\
\label{EWSBnew}
\end{eqnarray}

which we recast for later convenience in the following form:

\begin{eqnarray}
X_{m_3}^2( \tilde{\alpha} - X) t^4  + ( \tilde{\alpha} X_{m_1}^2 -
X (X_{m_1}^2 + X_{m_2}^2 ) ) t^3  && \nonumber\\ 
+ ( \tilde{\alpha} X_{m_2}^2 
+ X (X_{m_1}^2 + X_{m_2}^2 ) ) t +  X_{m_3}^2 ( \tilde{\alpha} + X)
=0&& \label{eqtbeta}\\
\end{eqnarray}
\begin{eqnarray}
u = \frac{1}{\tilde{\alpha} (t^2  -1)}
( X_{m_3}^2(t^2  + 1) + 
(X_{m_1}^2 + X_{m_2}^2 )t ) && \label{eqv1v2} \\
\nonumber
\end{eqnarray}
where $t\equiv \frac{v_2}{v_1} = \tan \beta$ and $u \equiv v_1 v_2$.

The $(8\times8)$ matrix of second derivatives 
$$\frac{\partial^2 V}{\partial H_a^i \partial H_b^j}|_{H_a=<H_a>}$$
has, (after having eliminated $X_{m_1}^2, X_{m_2}^2$ through
Eqs.(\ref{EWSBnew})), five non-zero eigenvalues and three vanishing ones
corresponding to the goldstone modes. It is technically easier 
and theoretically equivalent, to express
the stability conditions, Eq.(\ref{stabili}) in terms of some invariants
of the matrix, rather than in terms of the eigenvalues themselves.
One then finds the following conditions,

\begin{eqnarray}
&&-(v_1^2 + v_2^2) \frac{X_{m_3}^2}{v_1 v_2} \geq 0 \label{newinv1}
\end{eqnarray}
\begin{eqnarray}
&&2\tilde{\alpha} ( v_1^2 - v_2^2) + (v_1^2 + v_2^2)(2 X - \frac{X_{m_3}^2}{v_1 v_2} )\geq 0  \label{newinv2}
\end{eqnarray}
\begin{eqnarray}
&&-4 \tilde{\alpha}^2 v_1^2 v_2^2 + 2 (v_2^2 - v_1^2)[
(v_1^2 + v_2^2 ) \tilde{\alpha} - (v_2^2 - v_1^2) X] \frac{X_{m_3}^2}{v_1 v_2}
\geq 0 \label{newinv3}
\end{eqnarray}
\begin{eqnarray}
( -\frac{X_{m_3}^2}{v_1 v_2} + \tilde{\beta}) ( v_1^2 + v_2^2) \geq 0
\;\; \mbox{(twice)}
\label{newinv4}
\end{eqnarray}

We are now ready to discuss some features of the difference between tree-level
and one-loop minimization conditions. We note first that the only parameter 
appearing in Eq.(\ref{vloop}) and which is purely one-loop is $\tilde{\alpha}$.
In the tree-level limit $\tilde{\alpha} \to 0, X_{m_i}^2 \to m_i^2,
X \to g^2/8$ and $\tilde{\beta} \to g_2^2/2$ and Eqs.(\ref{EWSBnew}) yield
the usual tree-level minimization conditions. Furthermore, in this limit
Eqs.(\ref{newinv1} - \ref{newinv4}) become automatically satisfied once 
Eq.(\ref{newinv1}) is, that is when $ \frac{ m_3^2}{v_1 v_2} \leq 0$. However,
the previous inequality is itself a consequence of 
the tree-level version of Eqs.(\ref{EWSBnew}), as can be seen from solving 
Eq.(\ref{eqtbeta}) in terms of $t$ in the limit $\tilde{\alpha} \to 0$,
\cite{nous}. This explains the automatic positivity of the squared Higgs masses
stated in the introduction.\\
At one-loop level $\tilde{\alpha} \neq 0$. Even if $\tilde{\alpha}$ remains 
small\footnote{and the $X_{m_i}^2$'s,  $X$ and $\tilde{\beta}$ close to their
tree-level values so that the change in the effective potential at each point 
remains perturbatively small}, it leads to a drastic change in the structure of 
Eqs.(\ref{EWSBnew}) and (\ref{newinv1} - \ref{newinv4}). 
The stationarity equations can lead now to four different values for 
$\tan \beta$ (instead of two at tree-level), while the stability equations are 
not all satisfied when 
\begin{equation}
\frac{X_{m_3}^2}{v_1 v_2} \leq 0
\label{newcond1}
\end{equation}

Actually Eq.(\ref{newcond1}) guarantees Eqs.(\ref{newinv1}, \ref{newinv2}, 
\ref{newinv4}) while Eq.(\ref{newinv3}) will bring a new constraint.
Moreover, an important difference from the tree-level case is that
Eq.(\ref{newcond1}) is generically no more a consequence of the stationarity
conditions Eqs.(\ref{EWSBnew}). 

\section{The  $\tan \beta$ model-independent bounds}

In this section we present the new
constraints on $\tan \beta$ ensuing from Eqs.(\ref{EWSBnew})
and (\ref{newinv1} - \ref{newinv4}). We will give just a heuristic 
argument and make some general remarks concerning these constraints.
(The interested reader is referred to \cite{nous}
for full details of the study). To start with, we should stress that the 
solution $\tan \beta =1$ is from general considerations always unphysical
and should not be considered. Indeed it is easy to see from
Eqs.(\ref{EWSBnew}) that this solution would require 
\begin{equation}
X_{m_1}^2 + X_{m_2}^2 + 2 X_{m_3}^2 =0 \label{flatness}
\end{equation}
The above equation and $v_1=v_2$ imply that the effective potential
Eq(\ref{vloop}) is vanishing along this direction. This corresponds actually
to a D-flat direction together with a cancellation of the soft-susy terms
against some F-terms in Eq.(\ref{flatness}). In any case, the vanishing of the 
effective
potential means that all values of $v_1 (=v_2)$ are
degenerate. Thus $\tan \beta =1$ cannot lead to a preferred electroweak 
scale! The above is true at least up to one-loop order within our
approximation\footnote{D-flatness can easily suggest that it would also
be true to any order of perturbation theory, however as we said, one also
has to deal with the need to cancel the F-terms against the soft-terms}. 
Having barred this solution, Eq.(\ref{eqv1v2}) becomes well-defined. Actually
it is also well-defined in the limit $\tilde{\alpha} \to 0$ since in this limit
the numerator also goes to zero as can be seen from Eq.(\ref{eqtbeta}), so that
$u$ goes actually to a {\sl finite} computable quantity.\\
Let us now give a heuristic argument showing that the
interplay between Eqs.(\ref{eqv1v2}), (\ref{newcond1}) and (\ref{alpha})
forbids $\tan \beta$ to be too large or
too small. [We restrict ourselves throughout the discussion to 
positive $\tan \beta$. The fact that this choice does not reduce the generality
of the argument is somewhat tricky at the one-loop level \cite{nous}].
If the region $\tan \beta >>1$ were allowed, then Eq.(\ref{eqv1v2}) would 
behave like

\begin{equation}
 v_1 v_2 \sim \frac{X_{m_3}^2}{\tilde{\alpha}} 
\label{heur1}
\end{equation}
in that region. Taking into account Eq.(\ref{newcond1}) one then must have
$$ \tilde{\alpha} \leq 0$$
However, due to the form of $\tilde{\alpha}$ Eqs.(\ref{alpha},\ref{kapa})
and to the fact that 
$ \tan \beta \simeq \frac{Y_b}{Y_t} \frac{m_t}{m_b}$ one sees easily
that $ \tilde{\alpha} \leq 0$ and $\tan \beta$ arbitrarily large 
(i.e. $\frac{Y_t}{Y_b} >>1$) cannot
be simultaneously satisfied . We are thus lead to the conclusion that an
arbitrarily large $\tan \beta$ would give a contradiction, 
{\sl i.e. there should be a theoretical upper bound on $\tan \beta$}.\\
Similarly, in the region $|tan \beta| <<1 $ one has

\begin{equation} 
 v_1 v_2 \sim -\frac{X_{m_3}^2}{\tilde{\alpha}}
\label{heur2} 
\end{equation}
which, together with Eq.(\ref{newcond1}) implies that $\tilde{\alpha} \geq 0$,
the latter inequality being in contradiction with $|tan \beta| <<1 $.
Thus $\tan \beta$ cannot be arbitrarily small and {\sl a 
theoretical lower bound should exist}. To determine the actual
upper and lower bounds on $\tan \beta$ requires much more work. Here we only
state the results.

As we said in the previous section, there are only two independent constraints
from Eqs.(\ref{newinv1} - \ref{newinv4}), namely Eq.(\ref{newcond1})
and Eq.(\ref{newinv3}). The first one of these constraints combined with
Eq.(\ref{eqv1v2}) yields the following bounds,\\

{\bf a)}
\begin{eqnarray}
&\mbox{\, if \,} \tan \beta > 1& \;\;\; \mbox{then} \;\;\;\tan \beta_{-} \leq \tan \beta \leq
\tan \beta_{+} \\
&& \nonumber \\
&\mbox{where}& \;\;\; \tan \beta_{-}=Min( T_{+}, \frac{m_t}{m_b})  \nonumber\\
&\mbox{and}& \;\;\;\tan \beta_{+}=Max( T_{+}, \frac{m_t}{m_b}) \nonumber\\
\nonumber
\end{eqnarray}
\begin{eqnarray}
&\!\!\!\!\!\!\!\!\!\!\!\!\!\!\!\!\!\!\!\!\mbox{\, if \,} \tan \beta < 1& \;\;\; \mbox{then} \;\;\; T_{-} \leq \tan\beta <1 \nonumber \\
\end{eqnarray}
 
\begin{equation}
\mbox{\,where\,\,}
T_{\pm} = \frac{ -X_{m_1}^2 - X_{m_2}^2 \mp \sqrt{(X_{m_1}^2 + X_{m_2}^2)^2 -
4  X_{m_3}^4 }}{2 X_{m_3}^2} \nonumber
\end{equation}

while Eq.(\ref{newinv3}) can be re-expressed in the form\\

{\bf b)}
\begin{equation}
 \tan^2\beta \leq t_{-} \mbox{\,or\,} \tan^2\beta \geq t_{+}
\label{newcond2} 
\end{equation}

where 
\begin{equation}
t_{\pm}= \frac{ \tilde{\alpha}^2 \frac{v_1 v_2}{X_{m_3}^2} - X  \mp
\sqrt{( X - \tilde{\alpha}^2 \frac{v_1 v_2}{X_{m_3}^2})^2 + \tilde{\alpha}^2-
X^2 }}{ \tilde{\alpha} - X }
\label{newcond2p}
\end{equation}
leading to the second set of bounds. $t_{\pm}$ have an implicit dependence
on $\tan \beta$  through $v_1 v_2$, the latter being related to $\tan \beta$
through Eq.(\ref{eqv1v2}). 
[A preliminary version of these 
constraints was already given in \cite{moriond}, however, in a less general
form for {\bf a)} with no upper bound. It is the realization that the sign
of $\tilde{\alpha}$ places readily $\tan \beta$ on one side or the other
of $\frac{m_t}{m_b}$ which lead to the new form with the upper bound\footnote{
One should note that we rely here on the tree-level relations
$ m_t= \frac{Y_t v_2}{\sqrt{2}}, m_b = \frac{ Y_b v_1}{\sqrt{2}} $, which,
apart from leading log corrections in the running of $Y_t, Y_b, v_i$ ,
might suffer from some small corrections. Nonetheless, the qualitative form
of the constraint will not be altered}.]

Constraints {\bf a), b)} are necessary {\sl model-independent} conditions
for the existence of an electroweak (local) minimum. They can be readily 
implemented in any phenomenological analysis of the MSSM and allow to reduce
from the start, the allowed $\tan \beta$ domain. The actual analysis will
of course depend on the values of $t_{\pm} $ and $T_{\pm}$, these being 
calculable in terms of the other parameters of the MSSM. They have, however,
some general features:

\begin{eqnarray}
T_{-} \leq 1 \leq T_{+} \label{Tp}\\
t_{-} \leq 1 \leq t_{+} \label{tp}\\ 
\nonumber
\end{eqnarray} 
 
$T_{\pm}$ exist as far as the one-loop effective potential is bounded from
below, i.e. 
\begin{equation} 
X_{m_1}^2 + X_{m_2}^2 \pm 2 X_{m_3}^2 \geq 0
\end{equation}

while $t_{\pm}$ are effective all the time. In any case, Eq.(\ref{tp}) in 
conjunction with {\bf b)} shows that a region around $\tan \beta =1$ should
be always excluded. Furthermore, when $\tan \beta <1$ then {\bf a)} and
{\bf b)} imply

\begin{equation}
T_{-} \leq \tan \beta \leq t_{-}
\end{equation}
Thus this window will be theoretically closed if $t_{-} \sim T_{-}$.
This possibility of forbidding model-independently $\tan \beta < 1$
should be contrasted for instance with the usual argument in the context of 
SUGRA-GUT and requiring $b-\tau$ unification \cite{betau}.\\
On the other hand, when $\tan \beta >1$, {\bf a)} and {\bf b)} imply
\begin{equation}
Max( \sqrt{t_{+}}, \tan \beta_{-}) \leq \tan \beta \leq \tan \beta_{+}
\end{equation}

It is interesting to note here that when $T_{+} < \frac{m_t}{m_b}$
one has the model-independent bounds
\begin{equation}
1 \leq Max( \sqrt{t_{+}}, T_{+}) \leq \tan \beta \leq \frac{m_t}{m_b}
\label{better}
\end{equation}
These bounds are similar to those obtained in minimal sugra, namely, 
$ 1 < \tan \beta < \frac{m_t}{m_b} $, \cite{giudice}, but happen to have a more
restrictive lower bound! There is of course no contradiction here; it only 
means that using the tree-level (actually tree-level-RGE-improved) minimization
conditions together with some model-dependent assumptions like the universality
of the soft-susy masses at the unification scale, is weaker than just using
the model-independent
one-loop minimization conditions. Of course, one can still make the same
model-dependent assumptions on top of the one-loop minimization conditions
and get even stronger bounds than in Eq.(\ref{better}).\\
On the other hand, it is worth noting that one can enforce $\tan \beta$ 
to be greater or smaller than $\frac{m_t}{m_b}$ by choosing the free
parameters of the MSSM in such a way that $ T_{+} < \frac{m_t}{m_b}$ or
$> \frac{m_t}{m_b}$. In the second case one has,
\begin{equation}
\frac{m_t}{m_b} \leq \tan \beta < T_{+} 
\end{equation}
Again, all the above bounds should be contrasted with the more qualitative ones,
derived in the literature from the requirement of perturbativity of the 
Higgs-top or Higgs-bottom Yukawa couplings, \cite{habertalk}. It is clear that
the requirement of perturbativity is more of practical than theoretical
relevance, while our constraints are directly related to the physical
requirement of symmetry breaking. 

To summarize, we have established fully model-independent theoretical bounds
on $\tan \beta$ which are {\sl necessary} conditions for electroweak symmetry
breaking. They are actually a consequence of the conditions for (local) minima,
before imposing other physical requirements such as relating the vev's to
the correct electroweak scale. This shows that the structure of the MSSM,
even when freed from further model assumptions about the origin of electroweak
symmetry breaking, still leads to non-trivial constraints which should be taken
into account, both in fully model-independent or model-dependent
phenomenological analyses.   
 
\section{Outlook}
In this talk, we studied the structure of the (local) minima of the effective
potential only in the neutral Higgs fields directions. Other minima can occur
outside these directions and can lead to spontaneous breaking of the electric
charge and/or color symmetries \cite{casas}. 
The conditions for the occurrence of such
minima, or at least requiring a relatively short life-time of the associated
states, constitute important supplementary constraints. Further conditions
from the requirement of boundedness from below can also occur. However all 
these issues are usually addressed in the literature, overlooking the potential
modification the detailed structure of the one-loop corrections can bring
\cite{casas}. We illustrated in this talk an example of such effects.
An analytic study, along the same lines, is now being pursued in directions
involving squark and slepton fields.\\
Another issue concerns the gauge-fixing dependence of the results. This is safe
inasmuch as the results involve only stationary points of the effective
potential \cite{xsindep}. In this case, a local minimum remains so,
even if the magnitude of the curvature is gauge-fixing dependent.
However, in some circumstances, boundedness from below conditions are written 
at non-stationary points, in which case much more care is required. 
For instance one can show that such conditions can at best be necessary,
but certainly not sufficient, unless they are verified for a `large' class
of gauge-fixing terms or --which is more convincing-- if boundedness from
below is assured at the level of a more fundamental theory than the MSSM,
at the relevant very high energy scale.\\
Finally, it is important to take into account the detailed structure of the
logarithmic terms, keeping as much as possible the information simultaneously
from different scalar fields and mass scales. The main technical difficulty
in obtaining analytic information from the logs, is the need to diagonalize
the ``mass'' matrix $M^2$ in Eq.(\ref{EP}). To make this tractable, one often
resorts to some simplifying approximations, like keeping just the top-stop
and bottom-sbottom contributions to the effective potential. As these 
approximations are justified as far as one is concerned with the leading
loop corrections to tree-level observables 
(apart perhaps in some non-generic regions of the parameter space), it is less
clear that by doing so one does not overlook some theoretical constraints about the (local) minima, precisely in the field directions which are neglected.
This question is of particular relevance to the stability conditions,
since a straightforward analytic determination of the second order derivatives
in a sufficiently large scalar field space can quickly become intractable,
unless the field space is sufficiently reduced in order to diagonalize
$M^2$. But in this case a corresponding information is lost in the reduced
directions! The trick is to use a resummed formula which we derived for
the second order derivatives, valid {\sl before} reducing the field space. 
Then the field
space is reduced accordingly in order to obtain (analytically) 
diagonalizable matrices,
and in the same time keep the full information of the second order derivatives
in the initial field space. These issues are now under consideration.

{\bf Acknowledgments:} One of us (G.M.) would like to thank the organizers,
especially Argyris Nicolaidis, for the instructive and relaxed atmosphere
that prevailed during the joint meeting.

%\newpage

\end{document}